**Russian assimilatory palatalization is incomplete neutralization**

**Sejin Oh[1], Jason A. Shaw[2], Karthik Durvasula[3], Alexei Kochotov[4]**

[1] CNRS/Université Sorbonne-Nouvelle (se-jin.oh@cnrs.fr), [2] Yale University (jason.shaw@yale.edu), [3] Michigan State University (durvasul@msu.edu), [4] University of Toronto (al.kochetov@utoronto.ca)

## Abstract

Incomplete neutralization refers to phonetic traces of underlying contrasts in phonologically neutralizing contexts. The present study examines one such context: Russian assimilatory palatalization in C+j sequences. Russian contrasts plain and palatalized consonants, e.g., /p/ vs. /pʲ/ with the "plain" consonants having a secondary articulation, involving retraction of the tongue dorsum (velarization/uvularization). However, Russian also has stop-glide sequences that form near-minimal pairs with palatalized stops: e.g., /pjot/ 'drink (3ps pres)' vs. /pʲok/ 'bake (3ps past).' In the environment preceding palatal glides, the contrast between palatalized and plain consonants is neutralized, due to the palatalization of the plain stop: /pjot/→[pʲjot] (assimilatory palatalization). The purpose of the study is to explore whether the neutralization is complete. To do so, we conducted an electromagnetic articulography (EMA) experiment examining temporal coordination and the spatial position of the tongue body in underlyingly palatalized consonants and those derived from assimilatory palatalization. Articulatory results from four native speakers of Russian revealed that gestures in both conditions are coordinated as complex segments, i.e., they are palatalized consonants; however, there are differences across conditions consistent with the residual presence of a tongue dorsum retraction gesture in the "plain" obstruents. We conclude that neutralization of the plain-palatal contrast in Russian is incomplete—consonants in the assimilatory palatalization condition exhibit inter-gestural coordination characteristic of palatalized consonants along with residual evidence of an underlying tongue dorsum retraction (velarization/uvularization) gesture.

## 1. Introduction

Many instrumental investigations of phonologically neutralized contrasts have revealed that surface segments are not in fact identical to each other, as they contain phonetic traces of underlying contrasts. This phenomenon, often called incomplete neutralization, has been found for final devoicing in many languages. In the case of final devoicing, the voicing contrast is preserved in word-initial and word-medial positions. However, in the word-final position, both underlyingly voiced and underlyingly voiceless obstruents surface as voiceless. In German, for example, the voicing contrast of the alveolar stops in (1) is neutralized in word-final position, while being preserved in word-medial position as shown in (1).

(1) Examples of final devoicing in German (Adopted from Roettger et al., 2014)

| | |
|---|---|
| *Rat* /ʁa:**t**/ [ʁa:**t**] *'council'* | *Räte* /ʁæ:**t**ə/ [ʁæ:**t**ə] *'councils'* |
| *Rad* /ʁa:**d**/ [ʁa:**t**] *'wheel'* | *Räder* /ʁæ:**d**ɐ/ [ʁæ:**d**ɐ] *'wheels'* |

However, previous studies have provided considerable evidence that such phonological neutralization is phonetically incomplete in German (e.g., O'Dell & Port, 1983; Port & Crawford, 1989; Roettger et al., 2014), as well many other languages, such as Catalan (e.g., Charles-Luce & Dinnsen, 1987), Dutch (e.g., Warner et al., 2004), Polish (e.g., Slowiaczek & Dinnsen, 1985), and Russian (e.g., Dmitrieva et al., 2010; Kharlamov, 2012; 2014). Previous studies have shown that there are acoustic and articulatory differences between underlying voiced and voiceless obstruents, and such phonetic differences surface in the direction expected for the underlying form. More specifically, the underlyingly voiced obstruents tend to have shorter final stop closure durations, a shorter release burst, a longer preceding vowel, and/or more extensive voicing into closure than the underlying voiceless obstruents (Charles-Luce & Dinnsen, 1987; Dinnsen & Charles-Luce, 1984; Mascaró, 1987 for Catalan; Ernestus & Baayen, 2007; Warner et al., 2004 for Dutch; O'Dell & Port, 1983; Port & Crawford, 1989; Roettger et al., 2014 for German; Port & Crawford, 1989; Port & O'Dell, 1985; Slowiaczek & Dinnsen, 1985 for Polish; Dmitrieva et al., 2010; Kharlamov, 2012; 2014 for Russian).

Even though the majority of work on incomplete neutralization heavily focuses on final devoicing, the phenomenon is not restricted to this specific phenomenon. Other patterns that have long been described as neutralization have also turned out to be cases of incomplete neutralization: flapping in American English (e.g., Herd et al., 2010), coda aspiration in Eastern Andalusian Spanish (e.g., Gerfen, 2002), monomoraic lengthening in Japanese (e.g., Braver, 2019), vowel epenthesis in Levantine Arabic (e.g., Gouskova & Hall, 2009), vowel deletion in French (e.g., Fougeron & Steriade, 1997), blended vowels in Romanian (e.g., Marin, 2012), laryngeal neutralization in Korean (Lee, 2016), among others. There are also cases of what appears to be complete neutralization, such as Korean manner neutralization (Kim & Jongman, 1996; Lee, 2016).

Russian contrasts palatalized and plain consonants (so-called "soft" and "hard" consonants, respectively), as shown in (2) (e.g., Avanesov, 1972; Kochetov, 2002, 2006ab; Padgett, 2001; 2003; Timberlake, 2004). Plain consonants, however, get palatalized when followed by a palatal glide, leading to neutralization of the contrast in this particular context (e.g., Avanesov, 1972; Timberlake, 2004).[i] For example, the contrast between /$p^j$/ and /p/ (as in /$p^j$ok/ vs. /pjot/) is neutralized due to the palatalization of the plain stop in the consonant-glide sequence. Purely for expository convenience, we refer to the phonemic palatalized consonants as 'underlying palatalization' and to the plain consonants preceding a palatal glide (and thus surfacing as palatalized) as 'assimilatory palatalization'.

| (2) | Plain | Palatalized | Assimilatory palatalization |
|---|---|---|---|
| | /pot/ [pot] 'sweat' | /$p^j$ok/ [$p^j$ok] 'bake (3ps past)' | /pjot/ [$p^j$jot] 'drink (3ps pres)' |
| | /buk/ [buk] 'beech' | /$b^j$ust/ [$b^j$ust] 'bust' | /bjut/ [$b^j$jut] 'beat (3ps pl)' |

Interestingly, however, previous studies have reported that "plain" consonants in Russian may be characterized by a secondary articulation involving retraction of the tongue dorsum (velarization or uvularization; see Litvin, 2014; Padgett, 2001; 2003; Roon & Whalen, 2019; Skalozub, 1963). That is, the words /pot/ and /buk/ in (2) may in fact be /$p^{\gamma}ot^{\gamma}$/ and /$b^{\gamma}uk^{\gamma}$/. As Rubach (2000) argues, "there are no 'plain' consonants [in Russian, as] … every consonant is articulated with one of the following two tongue-body positions: forward movement and raising towards the hard palate (palatalization) or backward movement and raising towards the velum (velarization)" (p. 40). A similar view is taken by Padgett (2001). As phonological evidence for the underlying status of velarization, both authors cite the /i/-backing process, where this front vowel is consistently realized as central [ɨ] after 'plain' consonants (e.g., /igr-a-$t^j$/ 'to play (imperfective) vs. /s-igr-a-$t^j$/ 'to play (perfective)'). The backing of the vowel, the argument goes, is a natural assimilatory process triggered by an underlyingly velarized consonant: /$C^{\gamma}$-i/ $\rightarrow$ [Cɨ] (which is a near-mirror image of the assimilatory palatalization process examined in this study,

/Cˠj/ → [Cʲj]).

Skalozub (1963) is one of the early studies which systematically examined plain and palatalized consonants in Russian, using X-ray imaging, static palatography, odontography, and partial oscillography. Based on articulatory results from four Russian speakers, Skalozub (1963) argued that at least some plain consonants – lateral /l/ and labial consonants, are strongly velarized. Recent ultrasound studies by Litvin (2014) and Roon and Whalen (2019) further confirmed that plain consonants in Russian have a secondary articulation. Litvin (2014) examined plain fricatives and /l/ across different vowel contexts [a] and [ɛ]. Ultrasound data from six Russian speakers revealed that, regardless of the vowel context, /l/ and /f/ were uvularized, while /s/ and /ʂ/ were either velarized or uvularized. Roon and Whalen (2019) have also shown that plain consonants in Russian are velarized (and/or uvularized), subject to intra-speaker variation. In particular, their articulatory data from three Russian native speakers revealed that there were consistent and discernable dorsal gestures regardless of the manner and syllable position (initial vs. final), at least within labials /p/, /f/, and /m/, but the location of constriction varied by speaker (velar to uvular). Overall, similar conclusions about a secondary posterior gesture of Russian labials were reached in MRI studies by Kedrova et al. (2008; 4 speakers) and Biteeva (2021; a single speaker).

A question that arises from consideration of these patterns is whether the neutralization between plain and palatalized segments in Russian is phonetically (i.e., acoustically and/or articulatorily) complete. In other words, are there remaining traces of secondary velarization in words like [pʲjotˠ] and [bʲjutˠ], which are derived from /pˠjotˠ/ and /bˠjut/, respectively? To this end, the current study examines the phonetic realization of underlying and assimilatory palatalization, in which the underlying contrast between plain and palatalized consonants has been claimed to be phonologically neutralized.

The remainder of this paper is structured as follows: the rest of Section 1 provides background on Articulatory Phonology as well as past acoustic and kinematic studies on Russian palatalization. We then lay out our hypotheses and predictions in Section 2. In particular, we hypothesize that the gestural blending of two secondary articulation gestures (palatalization and velarization) would lead to incomplete neutralization of the underlying palatalization and assimilatory palatalization in Russian. Then, we transition to an empirical test of the hypotheses. We conducted an Electromagnetic Articulography (EMA) experiment examining temporal coordination and the spatial position of the tongue body for underlying and assimilatory palatalization. The methods of the experiment are described in Section 3, and the results are reported in Section 4. The discussion is presented in Section 5.

### *1.1. Articulatory Phonology*

Articulatory Phonology (henceforth, AP) provides a natural framework for describing incomplete neutralization. In this theory, the primitive phonological units are gestures. Gestures are discrete and abstract in the sense that they are specifically defined by a set of dynamical parameters which characterize each gesture distinctively (e.g., Browman & Goldstein, 1986; 1989; 1992; 1995; Pouplier, 2020). In AP, gestures are specified with respect to vocal tract variables. AP utilizes a set of gestural descriptors which distinguish contrastive gestures: Constriction degree (CD) and constriction location (CL). Tract variable goals (input values for CD and CL) determine the inherent spatial aspect. For example, /s/ and /ʃ/ differ in their values for CL (*alveolar* vs. *postalveolar*, respectively), while /s/ and /t/ differ in their values for CD (*critical* vs. *closed*, corresponding to fricative-like and stop-like constrictions). Finally, a dynamically control variable, stiffness (k), specifies the intrinsic temporal aspect of each gesture.

The spatiotemporal activation of gestures can be displayed in a gestural score with spatial information (specifications for tract variables) on the vertical axis and temporal information on the horizontal axis (e.g., Browman & Goldstein, 1989; 1992). For example, a gestural score for the word ‘pen’ /pɛn/ would include the input values for CD and CL of each gesture as well as their intergestural timing. Specifically, there are gestures associated with the initial /p/: a lip closure gesture and a wide glottal gesture. The tongue body gesture for /ɛ/ also starts at the beginning of the utterance overlapping with the gestures associated with /p/. The final consonant /n/ also has two gestures: a tongue tip closure and a velic opening, which also overlap with the preceding vowel gesture. The overlap between the velic opening and the vowel gesture leads to partial nasalization of the vowel.

Browman and Goldstein (1989) proposed that phonological phenomena such as deletion, insertion, assimilation, and weakening can be captured by two general processes: ‘hiding’ and ‘blending’ of gestures. When gestures significantly overlap on the different articulatory tiers, one gesture may hide the other acoustically, despite both gestures still being present articulatorily. For example, the apparent deletion of /t/ in ‘perfect memory’ ([ˈpʰɚfəkt ˈmɛməɹi]) at a fast speech rate is better described as gestural hiding (Tiede et al., 2001): the alveolar gesture for /t/ completely overlaps with the preceding velar gesture for /k/ and the following labial gesture for /m/, resulting in the former consonant being acoustically hidden. On the other hand, when two gestures overlap on the same articulatory tier, they compete with each other to achieve their own articulatory targets. This kind of overlap may lead to ‘blending’ of the dynamical parameters of these gestures. The gestural outcome of blending is different from that of either of the individual gestures. Instead, the outcome falls somewhere in between the two gestures, the extent of which depends on the strength of each gesture. For example, as discussed earlier, the backing of /i/ to [ɨ] after plain consonants in Russian is better described as a gestural blending between /i/ and the velarization gesture of the preceding consonant. In particular, the gesture for /i/ and /ˠ/ overlap on the same TB tract variable, resulting in the blending of CD and CL parameters for both /i/ (narrow, palatal) and /ˠ/ (critical, velar). In this blending process, the gesture with stronger blending parameters (e.g., the gesture for /i/) has a stronger influence on the output, while the output is still mildly affected by the gesture with weaker blending parameters (e.g., the velarization gesture).

In AP, incomplete neutralization does not require any special machinery. It follows from the blending of two gestures such that one gesture dominates control of the articulator but the other still has some influence.

### *1.2. Past results on Russian palatalization*

Independent of whether the contrast between underlying and assimilatory palatalization is neutralized or not, the consonant-glide sequence itself is not necessarily identical to the palatalized consonant. In fact, previous studies reported that there is a perceivable difference between palatalized consonants ($C^j$) and consonant-glide sequences (Cj). For example, Ladefoged and Maddieson (1998, p. 364) showed spectrograms comparing the Russian initial palatalized labial stop /pʲ/ and a labial stop + palatal glide sequence /pj/ (/pʲotr/ ‘Pyotr (name)’ vs. /pjot/ ‘drink (3ps pres)’). They observed that for the former F2 began falling immediately after the consonant release, while for the latter the F2 decrease began later, after a steady-state period.

In a more extensive study of the contrast, Diehm (1998) examined acoustic characteristics of various palatalized consonants ($C^j$) and corresponding consonant-glide sequences (Cj) produced by native speakers and learners of Russian. Results from eight native speakers (4 male and 4 female) revealed that consonant-glide sequences (Cj) exhibited significantly higher F2 at the transition onset than palatalized consonants ($C^j$) (2704 Hz vs. 2362 Hz for females; 2233 Hz vs. 2012 Hz for males). In addition, she reported that consonant-glide sequences (Cj) showed a significantly longer F2 steady-state duration than palatalized consonants ($C^j$) (on average 117 ms vs. 33 ms for females; 102 ms vs. 25 ms for males).

In addition, Suh and Hwang (2016) examined palatalized consonants ($C^j$) and consonant-glide sequences (Cj) in Russian, comparing them to a palatal glide in Korean. To measure glide duration, they first examined the vocalic duration comprising the j+V portion (from the onset of the vocoid to the offset of the vowel). They further calculated the durational ratio of the j+V portion to the pure vowel duration in CV. The results from five Russian native speakers revealed that the vocalic duration comprising the j+V portion of CjV syllables was significantly longer than the j+V portion of $C^j$V syllables.[ii]

Articulatory studies of Russian have shown differences that are consistent with the observations from acoustic data. Kochetov (2006b) examined the effect of syllable position on gestural organization using kinematic data from EMMA (Electromagnetic Midsagittal Articulometer). In particular, he compared articulatory patterns exhibited by the palatalized stop /$p^j$/, the plain stop /p/, and the palatal glide /j/ produced by four native speakers of Russian. The results revealed that the palatal gesture was longer when it occurred as a segment in /p#j/ sequences than when it occurred as secondary palatalization in /$p^j$/. In addition, and of particular interest to the present study, Kochetov showed that the relative timing of the labial gesture and the palatal glide gesture in stop-glide sequences (/p#j/) differed from the relative timing of these gestures in palatalized stops like /$p^j$/. More specifically, the glide gesture was achieved later in the stop-glide sequence /p#j/ than in the glide gesture for the palatalized stop /$p^j$/. However, since the stop-glide sequence used in the study involved a word boundary, it is unclear whether the delayed glide gesture in the segment sequence was due to the characteristics of the segment sequence or from the effect that the prosodic boundary may have had on articulatory timing. Consequently, the difference in the delayed achievement lag for /p#j/ and /$p^j$/ is not a valid criterion for accessing incomplete neutralization of underlying and assimilatory palatalization in Russian, nor is it a valid criterion for distinguishing complex segments and segment sequences more generally.

The acoustic and articulatory results summarized above confirm that there are phonetic cues to the difference between palatalized consonants ($C^j$) and consonant-glide sequences (Cj). These differences likely reflect the structural difference between a glide gesture as a secondary articulation and a glide gesture as a separate segment. Crucially, however, the acoustic differences between consonant-glide sequences (Cj) and palatalized consonants ($C^j$) do not provide any information as to whether the "plain" consonant in the consonant-glide sequences is palatalized or not. To evaluate incomplete neutralization, it is first necessary to establish whether the consonant preceding a palatal glide is indeed palatalized.

Such a quantification of palatalization in Russian might be achieved by examining temporal coordination for complex segments and segment sequences, as proposed by Shaw et al. (2021). The authors hypothesized that complex segments have a temporal basis—two articulatory gestures, G1 and G2, belong to the same complex segment if the onset of G2 is temporally coordinated with the onset of G1. In contrast, two gestures belong to sequences of segments if the onset of G2 is temporally coordinated with the offset of G1. These competing coordination relations were explored by investigating how the lag between the onset of G1 and the onset of G2 varied with G1 duration. The key finding involved differences between English consonant-glide sequences [pj], [bj], [mj], and [vj] (e.g., *pew*, *butte*, *muse*, *view*), and Russian palatalized labials [$p^j$], [$b^j$], [$m^j$], [$f^j$], [$v^j$]. Articulatory results from four English native speakers and four Russian native speakers revealed that for English stop-glide sequences, as consonant duration increased, so too did the lag between consonant and glide gestures (see Figure 1). In contrast, for Russian palatalized consonants, variation in duration had no effect on lag. That is, English stop-glide sequences showed the hypothesized temporal basis for segment sequences, while the palatalized consonants in Russian exhibited the hypothesized temporal basis for complex segments. The pattern of covariation successfully differentiated between palatalized labials in Russian, cases of underlying palatalization, and labial-glide sequences in English. In this paper, we examine whether this pattern of covariation differentiates two types of Russian sequences: underlying palatalization vs. assimilatory palatalization.

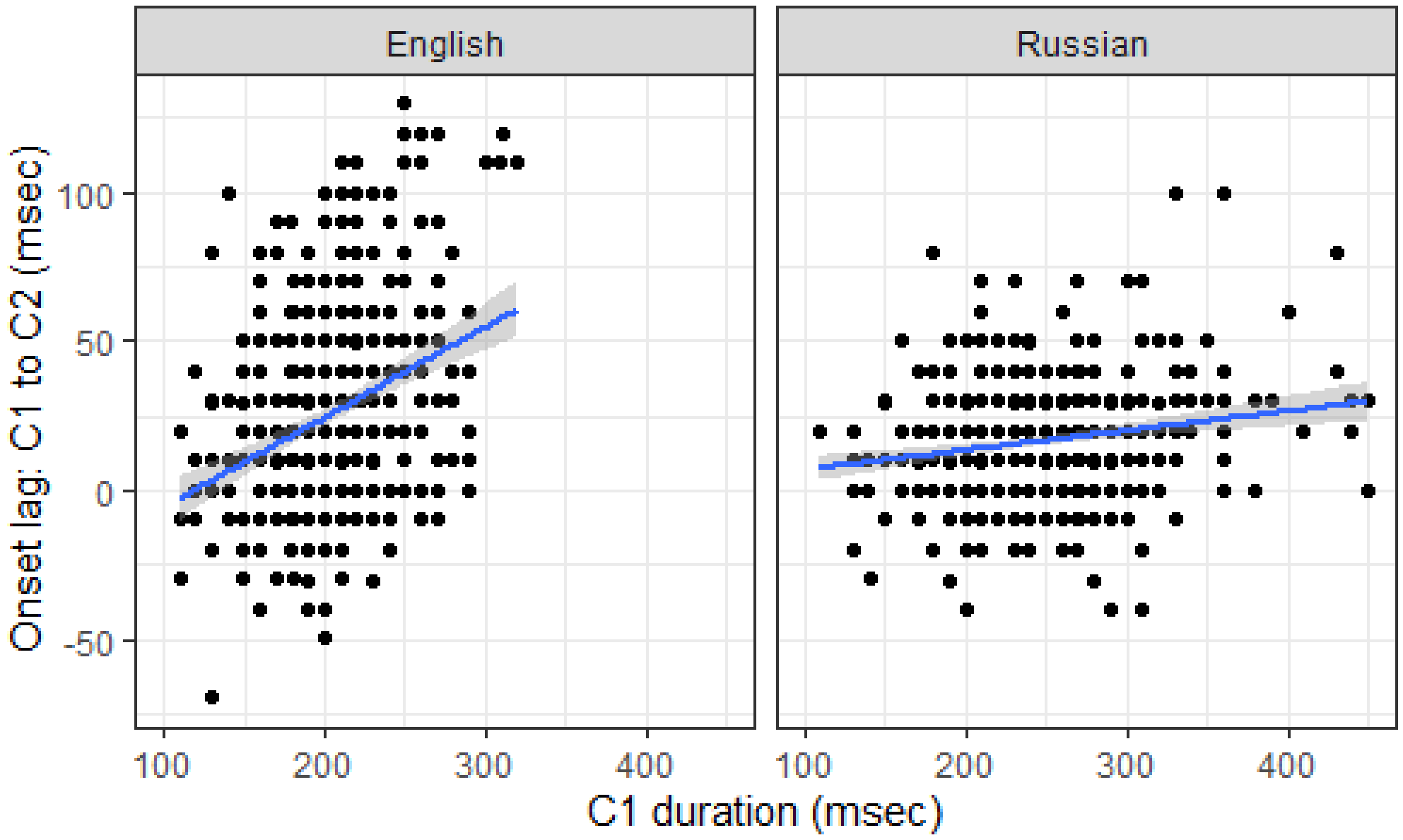

**Figure 1: A scatter plot of the effect of G1 duration (x-axis) on onset-to-onset lag (y-axis) for each language. English (a), which parses the gestures into segment sequences, shows a strong positive correlation while Russian (b), which parses the gesture into complex segments, shows no correlation (adapted from Shaw et al., 2021, p. 464).**

In the next section, we present our hypotheses regarding incomplete neutralization of underlying and assimilatory palatalization in Russian, making use of the temporal diagnostics for complex segmenthood.

## 2. Research questions and Predictions

The fundamental question of this study is whether two cases of Russian palatalization represent a case of incomplete neutralization. We have divided this into two sub-questions as follows:

- Research question 1: Do underlying palatalization (e.g., /b$^j$/) and assimilatory palatalization (e.g., /bj/ [b$^j$j]) both exhibit the temporal coordination characteristic of complex segments?
- Research question 2: Do underlying palatalization (e.g., /b$^j$/) and assimilatory palatalization (e.g., /bj/ [b$^j$j]) exhibit any systematic spatial and/or temporal differences?

The first research question addresses whether the two cases under study are indeed both cases of Russian palatalization, i.e., resulting in complex segments. If plain consonants preceding a palatal glide (assimilatory palatalization) are palatalized, this results in neutralization of underlying and assimilatory palatalization in Russian. We examine neutralization using the temporal diagnostics of complex segments and segment sequences. That is, if Russian palatalization exhibits neutralization, both underlying and assimilatory palatalization will show the temporal coordination of complex segments. In contrast, if Russian palatalization exhibits no neutralization, the underlying palatalization will show the temporal coordination of complex segments, while the assimilatory palatalization will exhibit the temporal coordination of segment sequences.

The second research question addresses whether the neutralization is complete (if the neutralization exists). That is, if there are spatial and/or temporal differences between the underlying and assimilatory palatalization, it would be considered incomplete neutralization. Assuming that plain consonants also have secondary velarization, we examine the completeness of the neutralization using the spatial position of the tongue body, as well as the temporal lag between the onset of the labial gesture and the onset of the palatal gesture.

Consequently, there are three possible outcomes depending on the temporal organization and spatial and/or temporal differences of underlying and assimilatory palatalization: no neutralization, complete neutralization, and incomplete neutralization.

- No neutralization: Underlying palatalization shows temporal coordination of complex segments, while assimilatory palatalization exhibits temporal coordination of segment sequences. Consequently, there are consistent spatial and/or temporal differences of underlying and assimilatory palatalization.
- Complete neutralization: Both underlying and assimilatory palatalization cases show the temporal coordination of complex segments, and there are no spatial and/or temporal differences between underlying and assimilatory palatalization.
- Incomplete neutralization: Both underlying and assimilatory palatalization cases show the temporal coordination of complex segments, and yet there are phonetic traces —spatial and/or temporal differences— indicative of the underlying categories, i.e., a palatal gesture for underlying palatalization and a velar/uvular gesture for plain consonants.

Given that plain consonants have secondary velarization (Litvin, 2014; Roon & Whalen, 2019; Skalozub, 1963), we predict that the gestural blending of two secondary articulation gestures (palatalization and velarization/uvularization) in assimilatory palatalization will lead to incomplete neutralization of underlying and assimilatory palatalization in Russian.

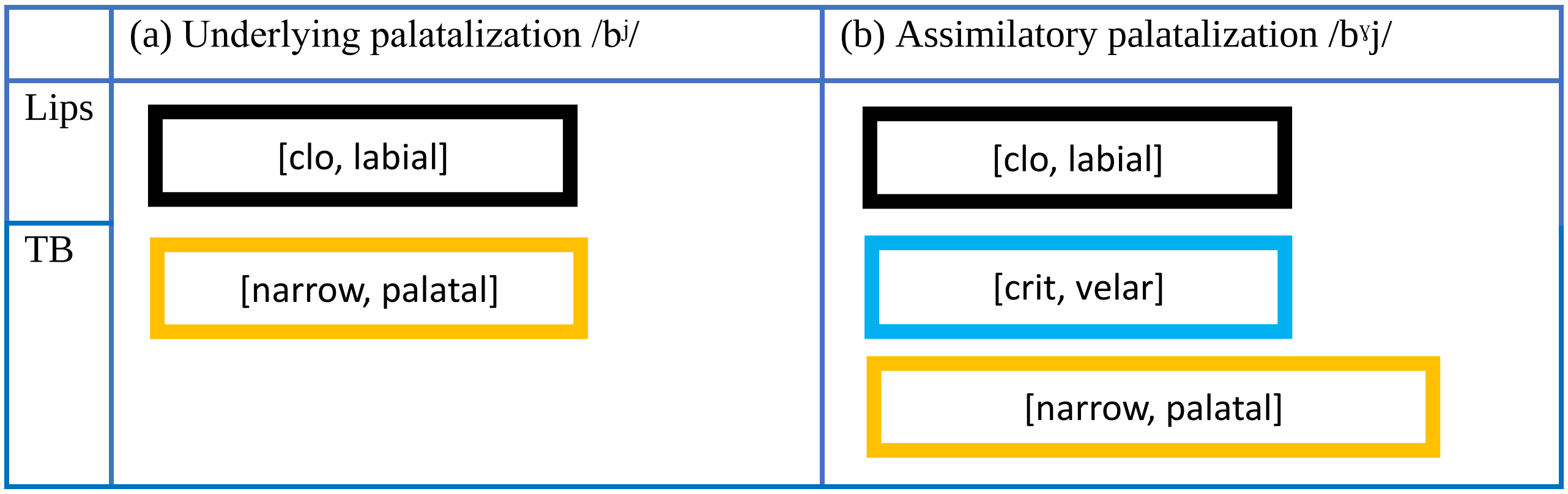

**Figure 2: Predicted gestural scores for underlying (a) and assimilatory palatalization (b) in Russian (incomplete neutralization).**

Figure 2 illustrates the hypothesis that motivates our predictions. It shows hypothesized gestural scores for underlying and assimilatory palatalization in Russian (see Section 1.1. for background on gestural scores). For underlying palatalization (panel a), there is a labial gesture and a palatal gesture, which start at the same time; for assimilatory palatalization (panel b) there is additionally a velar gesture overlapping in time with the labial and palatal gestures. Gestural overlap on the same tract variable, in this case, palatalization and velarization on the TB tract variable, would lead to gestural blending between these two gestures. Depending on the language-specific gestural blending parameters (see for discussion, e.g., Iskarous et al., 2012), blending could result in a slightly more retracted tongue position for assimilatory palatalization compared to underlying palatalization, which only has the palatal gesture on the TB tract variable. Consequently, this difference would lead to incomplete neutralization between underlying and assimilatory palatalization in Russian.

## 3. Methods

### *3.1. Participants*

Four native speakers of Russian participated in this experiment (3 female and 1 male). All speakers were in their 20s at the time of recording and living in the United States. The Russian speakers were born in Russia and moved to the United States as adults.

### *3.2. Materials*

The materials included six closely matched pairs representing two conditions: palatalized consonants vs. plain consonants preceding a palatal glide (**UNDERLYING** vs. **ASSIMILATORY PALATALIZATION**). In all cases, the primary word stress falls on the first syllable, and the vowel immediately following is either /u/ or /o/, as shown in Table 1. The target words were read in the carrier phase: *Она* ___ *повторила* /oˈn-a ___ po-vtoˈr-i-l-a / [ʌˈna ___ pəftʌˈrʲilʌ]. 'She ___ repeated.'

**Table 1: Russian target words**

| **Palatalized consonants (UNDERLYING palatalization)** | | | | **Consonant-glide sequences (ASSIMILATORY palatalization)** | | | |
|---|---|---|---|---|---|---|---|
| **word** | **IPA** | | **gloss** | **word** | **IPA** | | **gloss** |
| *пёк* | /pʲokˠ/ | [pʲokˠ] | bake (3ps past) | *пьёт* | /pˠj-o-tˠ/[iii] | [pʲjotˠ] | drink (3ps pres) |
| *бюст* | /bʲusˠtˠ/ | [bʲusˠtˠ] | bust (breast) | *бьют* | /bˠj-u-tˠ/ | [bʲjutˠ] | beat (3pp pres) |
| *мю* | /mʲu/ | [mʲu] | Mu (μ) | *Мью* | /mˠju/ | [mʲju] | a Pokémon name |
| *Фёдор* | /ˈfʲodˠorˠ/ | [ˈfʲodˠʌrˠ] | Fyodor (name) | *фьорд* | /fˠjorˠdˠ/ | [fʲjorˠd̥ˠ] | fjord |
| *вёз* | /vʲozˠ/ | [vʲoz̥ˠ] | carry (3ps past) | *вьёшь* | /vˠj-o-ʃˠ/ | [vʲjoʃˠ] | weave (2ps pres) |
| *вёдра* | /ˈvʲodˠrˠ-a/ | [ˈvʲodˠrˠʌ] | bucket (pl) | *вьёт* | /ˈvˠj-o-tˠ-sʲa/ | [ˈvʲjotsʌ] | weave (3ps pres refl) |

### *3.3. Procedure*

Data collection took place in the Phonetics Lab at Yale University Department of Linguistics. The articulatory and acoustic data were simultaneously recorded by means of *5D Electromagnetic Articulography (EMA)* and an audio-recording setup. To collect articulatory data, 9 sensors were attached to the participants: 3 sensors for tongue movements, 2 for lip movements (upper and lower lips), 1 for jaw movements (lower incisor), and 3 for reference (the nasion and left/right mastoids). Three sensors on the tongue, tongue tip (TT), tongue blade (TB), and tongue dorsum (TD), were attached along the sagittal midline of the tongue, being placed behind the tongue tip approximately 1 cm, 3 cm, 5 cm, respectively. Sensors were tracked using the *NDI Wave Speech Production System*. Reference sensors were used to computationally correct for head movements. As a post-processing procedure, the data was computationally corrected for head movements and rotated to the occlusal plane so that the bite of the teeth serves as the origin of the spatial coordinates. We also calculated a lip aperture trajectory, as the Euclidean distance between the upper and lower lip sensors. All trajectories were smoothed using the robust method described in Garcia (2010).

### *3.4. Analysis*

The post-processed data was visualized in *MVIEW* (Tiede, 2005). Changes in Lip Aperture (LA), computed as the Euclidean distance between the upper and lower lip sensors over time, were used to identify labial gestures. The TB sensor indexed the palatal gesture. Gestural landmarks were parsed with reference to the velocity signal using the *findgest* function in MVIEW. Specifically, the gesture *Onset* and *Target* landmarks were labeled at 20% of peak velocity in the movement toward constriction (see Figure 3). *Release* and *Offset* landmarks were labeled at a 20% threshold of peak velocity in the movement away from constriction. As illustrated in Figure 4, the two key temporal intervals computed from these articulatory landmarks were (1) *$G_1$ duration*, defined as the interval from *Onset* to *Offset* of the labial gesture; and *onset-to-onset lag*, defined as the interval between the *Onset* of the labial gesture (*$G_1$*) and the *Onset* of the palatal gesture (*$G_2$*). In addition to temporal coordination, the current study measured the longitudinal position (front-back) of the TB sensors at palatal gesture onset to assess any impact of underlying velarization on the realization of assimilatory palatalization. The spatial position of the TB sensors was normalized using z-scores for each speaker. Before proceeding with statistical analysis, we removed outliers that were greater than three standard deviations from the speaker-specific mean value of either *$G_1$ duration*, 7 tokens removed (0.6% of the data), or *onset-to-onset lag*, 18 tokens removed (1.6% of the data).

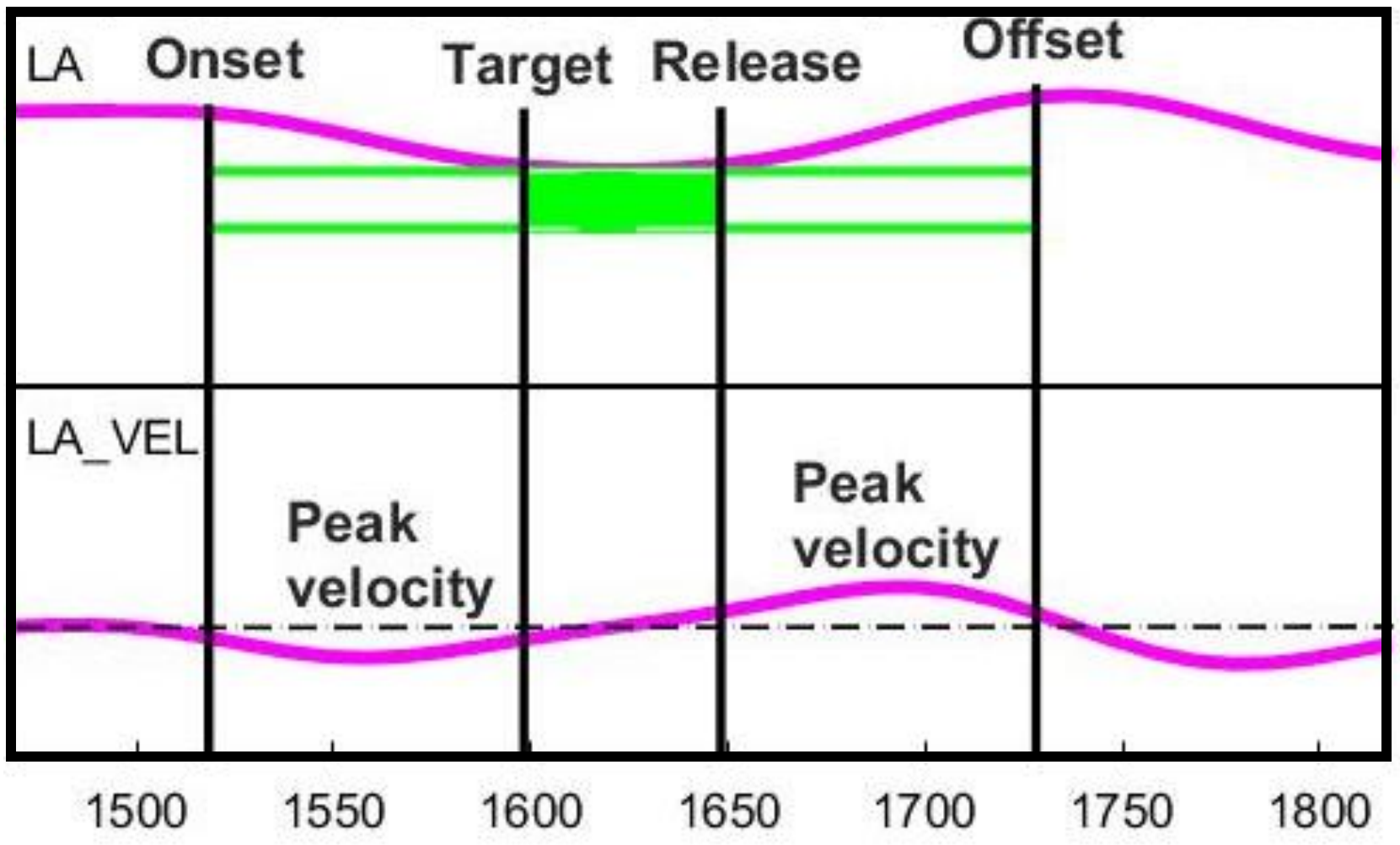

**Figure 3: Example of gesture parse for a labial gesture. The gestural landmarks, Onset, Target, Release, Offset, are labeled at 20% thresholds of peak velocity.**

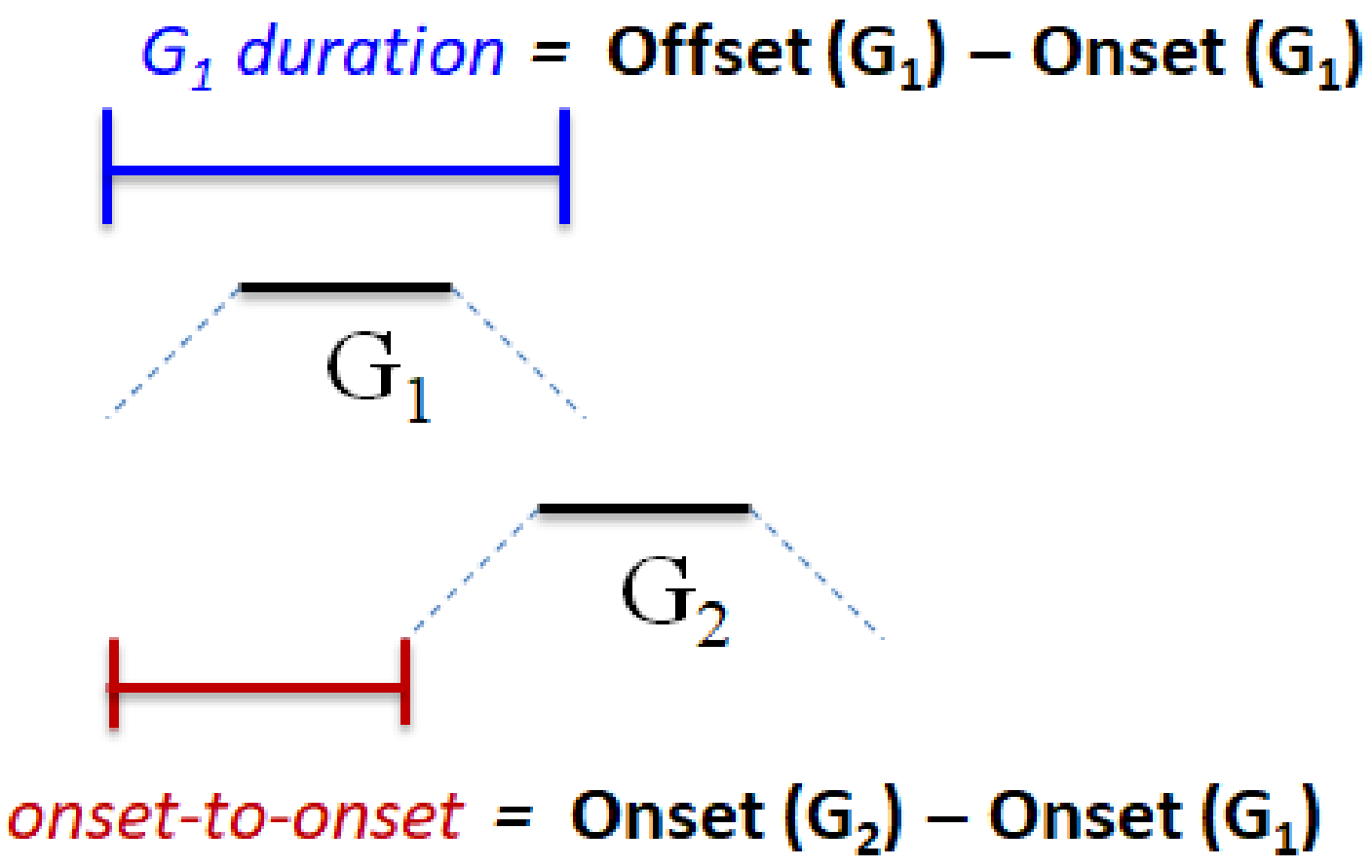

**Figure 4: Schematic depiction of the two intervals, G1 duration and onset-to-onset lag. G1 refers to the labial gesture and G2 refers to the palatal gesture.**

To examine the neutralization of Russian palatalization, the correlation between *onset-to-onset lag* and $G_1$ *duration* was analyzed. As $G_1$ *duration* varies, we ask whether *onset-to-onset lag* will positively covary, or whether these intervals will be statistically independent. If the contrast between a palatalized consonant (underlying palatalization) and a plain consonant preceding a palatal glide (assimilatory palatalization) in Russian is preserved (no neutralization), underlying palatalization will show no correlation between consonant duration and *onset-to-onset lag*, while for assimilatory palatalization, *onset-to-onset lag* will increase with $G_1$ *duration*, leading to a positive correlation between them. However, if the contrast is neutralized, both the underlying and assimilatory palatalization will exhibit no correlation between $G_1$ *duration* and *onset-to-onset lag*.

We, therefore, treat *onset-to-onset lag* as a dependent variable and evaluate whether *G1 duration* and *Condition* are significant predictors. We fit linear mixed-effects models to onset-to-onset lag using the *lme4* package in R (Bates et al., 2014). To a baseline model, consisting of by-subject and by-time random slopes for G1 duration and random intercepts for subjects and items, we added fixed factors of interest incrementally. First, we added *$G_1$ duration*, then *Condition* (UNDERLYING vs. ASSIMILATORY, with UNDERLYING as the reference level), and finally the interaction between *$G_1$ duration* and *Condition*. This gives a set of four nested linear mixed-effects models. We evaluated the significance of each fixed factor through *anova*. The fixed factor of primary interest is the interaction term: *$G_1$ duration* X *Condition*. If the contrast is not neutralized, *$G_1$ duration* is predicted to have a positive influence on *onset-to-onset lag* for assimilatory palatalization but not for underlying palatalization. On the other hand, if the contrast is neutralized, both palatalizations will exhibit the same pattern, i.e., the *$G_1$ duration* X *Condition* interaction will not be significant and there will be no correlation between *$G_1$ duration* and *onset-to-onset lag*.

To assess the incompleteness of the neutralization, we also examined the effect of *Condition* on the *TB position* at palatal gesture onset. If the neutralization is complete, there will be no difference in the *TB position* across conditions. However, if the neutralization is incomplete, the assimilatory palatalization will exhibit a more retracted tongue position than the underlying palatalization. To test this, separate linear mixed-effects models were run with *TB position* as a dependent variable and *Condition* as a fixed factor. *Speaker* and *Item* were included in all models as random intercepts

## 4. Results

### *4.1. Kinematic trajectories and distribution*

We first examine the continuous kinematic trajectories of relevant articulators for **UNDERLYING** and **ASSIMILATORY** palatalization in Russian. Figure 5 (a) illustrates kinematic trajectories for the item /b$^j$ust/ (**UNDERLYING** palatalization), as produced by the four Russian speakers in the study. The figure plots the LA trajectory in the upper panels and the TB trajectory in the lower panels. Each trajectory is represented by a different color; the thick dotted line shows the average trajectory. The temporal window of the trajectories is 600 ms long, spanning from 100 ms before the onset landmark of the lip aperture gesture to 500 ms following this landmark. The level of variability in the magnitude of the gestures varies by subject. For R2, most tokens occur tightly clustered around the mean; R1 and R3 show more variability, and R4 shows even more. On the other hand, the relative timing of the gestures appears similar across speakers - the fall in the LA trajectory, indicating the closing of the lips tends to coincide with the rise of the TB for the palatal gesture. To facilitate comparison, vertical gray lines indicate when the LA trajectory starts to fall (based on the average trajectory) and when TB starts to rise (also based on the average).

Figure 5 (b) shows kinematic trajectories for the item /b$^ɣ$jut$^ɣ$/ (**ASSIMILATORY** palatalization). The level of variability in the magnitude of the gestures appears similar to the **UNDERLYING** palatalization case. Regarding the relative timing of the gestures for the token /b$^ɣ$jut$^ɣ$/, the rise for the TB movement tends to follow shortly after the fall of the LA trajectory.

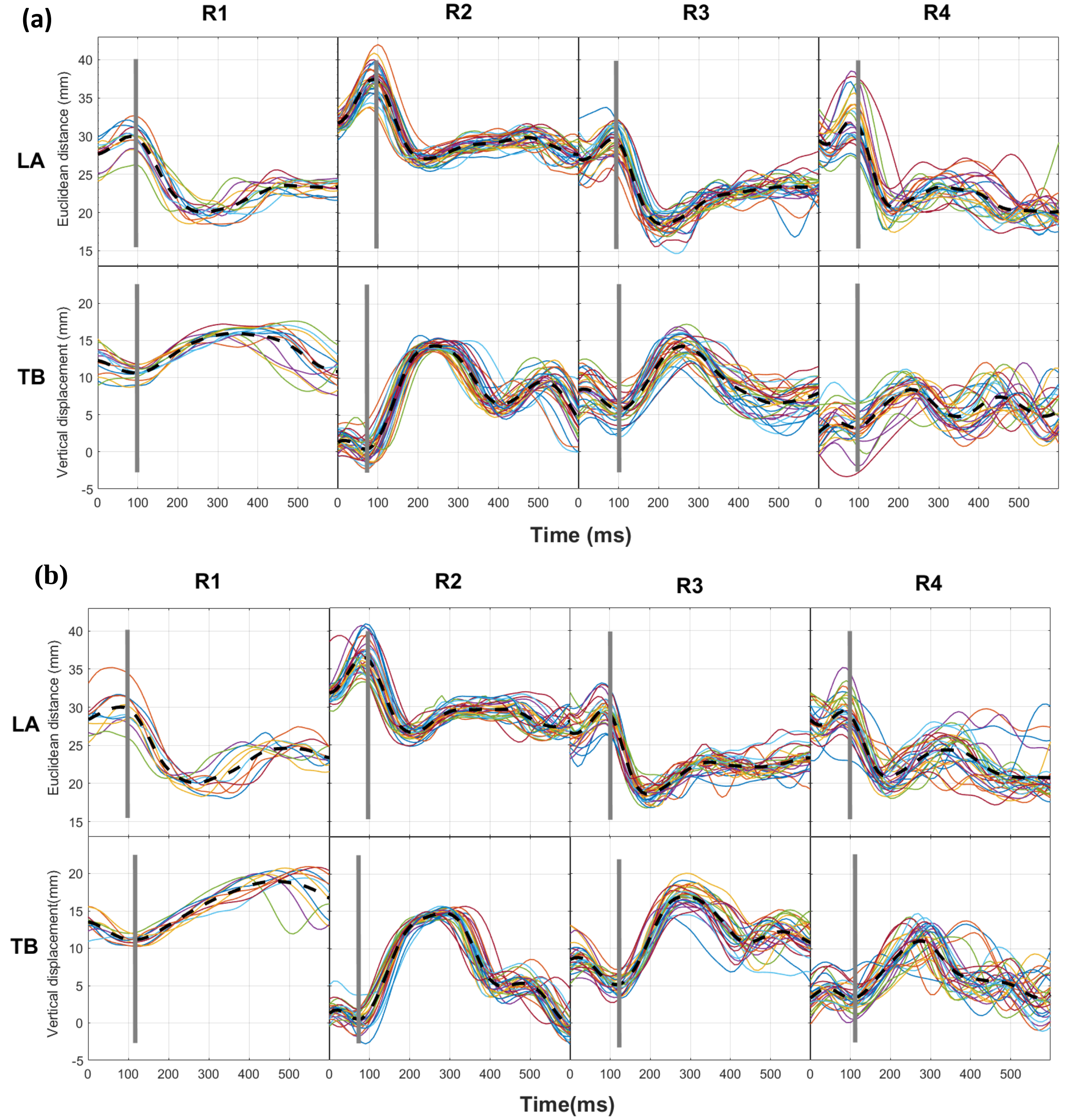

**Figure 5: (a) Tokens of /bʲust/ (UNDERLYING palatalization); (b) Tokens of /bˠjut/ (ASSIMILATORY palatalization). The thick dashed black line represents the average trajectory for each speaker. The top panels show the lip aperture (LA) trajectory. The bottom panels show the tongue blade (TB) in the vertical dimension. The time window of 600 ms extends from 100 ms before the onset of lip aperture movement to 500 ms after the onset of lip aperture movement. The vertical grey lines indicate the onset of LA lowering and the onset of TB raising, both based on the average trajectory.**

Next, we present the distribution of the continuous variables, the key intervals for the temporal coordination analysis: $G_1$ duration (labial gesture; See Figure 6) and onset-to-onset lag (Figure 8). For both intervals, we present the distribution by *Condition*: **UNDERLYING** vs. **ASSIMILATORY** palatalization. Also, for completeness, we plot the distribution of $G_2$ duration (palatal gesture; See Figure 7). This measurement is not directly related to our research questions, but is included for reference.

As shown in Figure 6, the $G_1$ duration measures have a slightly right-skewed distribution with a long right tail, which is common for temporal measurements of speech associated with linguistic units. This is true for the distributions of palatal gesture duration as well as onset-to-onset lag. Notably, however, the distributions of $G_1$ duration for **UNDERLYING** and **ASSIMILATORY** palatalization are heavily overlapped, with similar means and variance. On the other hand, as shown in Figure 7, **ASSIMILATORY** palatalization tends to have a longer palatal gesture than **UNDERLYING** palatalization, consistent with the previous findings (Kochetov, 2006b). The distribution of onset-to-onset lag shows that **ASSIMILATORY** palatalization tends to have a longer onset-to-onset lag than **UNDERLYING** palatalization and the distributions differ in shape, with **UNDERLYING** palatalization having a sharp peak with more values close to the mean (see Figure 8).

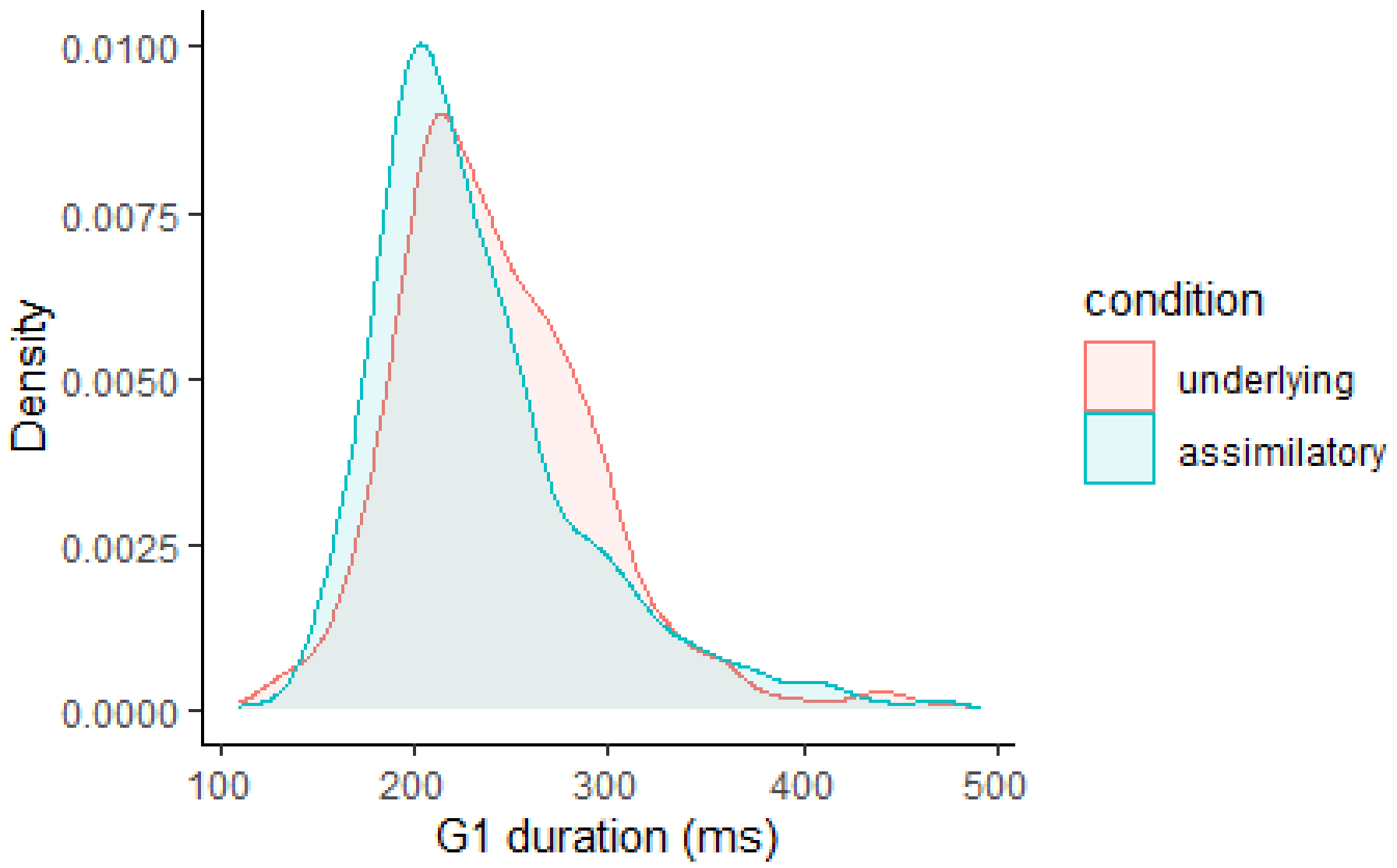

**Figure 6: The distribution of G1 (labial consonant) duration by *Condition*.**

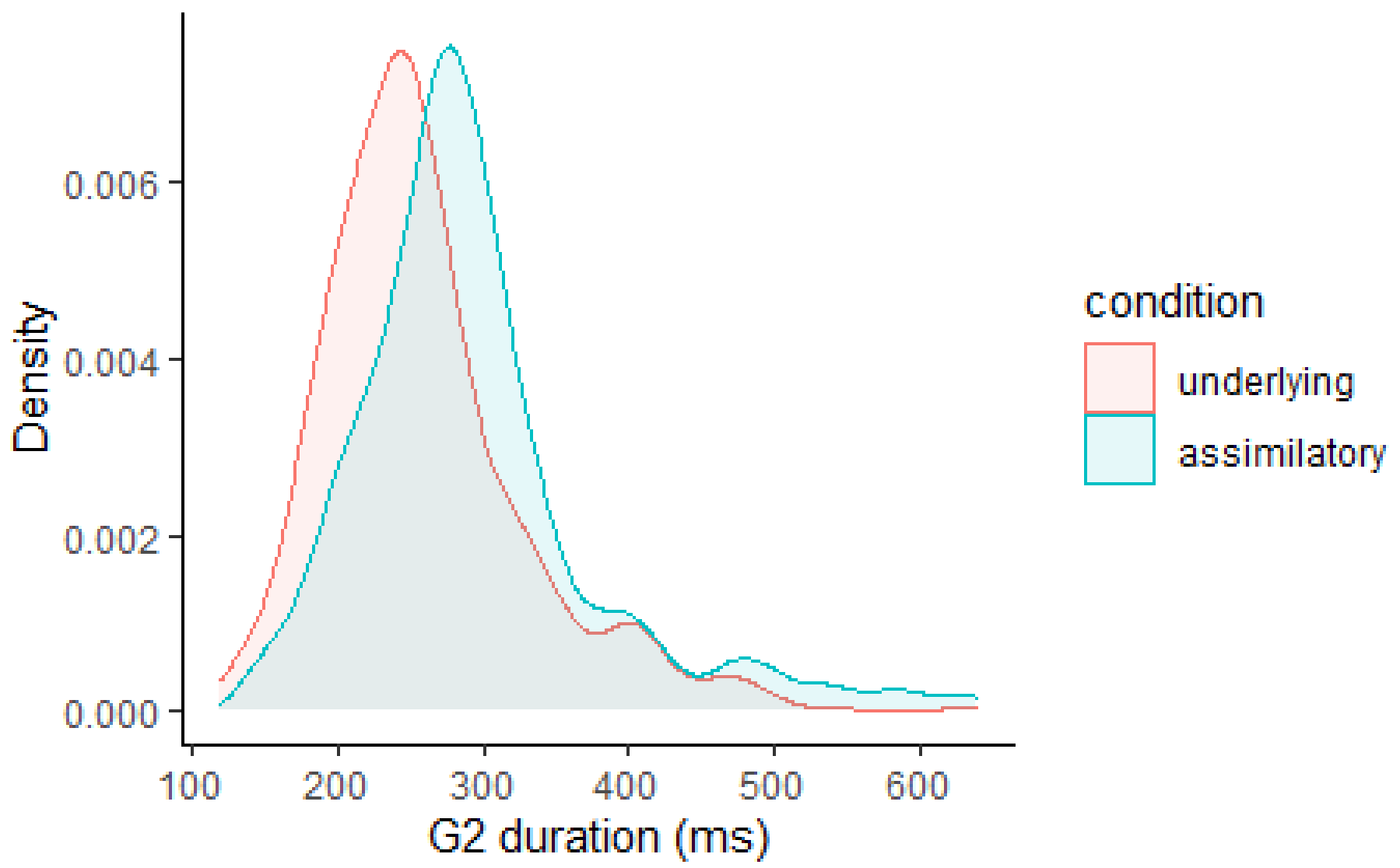

**Figure 7: The distribution of G2 (palatal gesture) duration by *Condition*.**

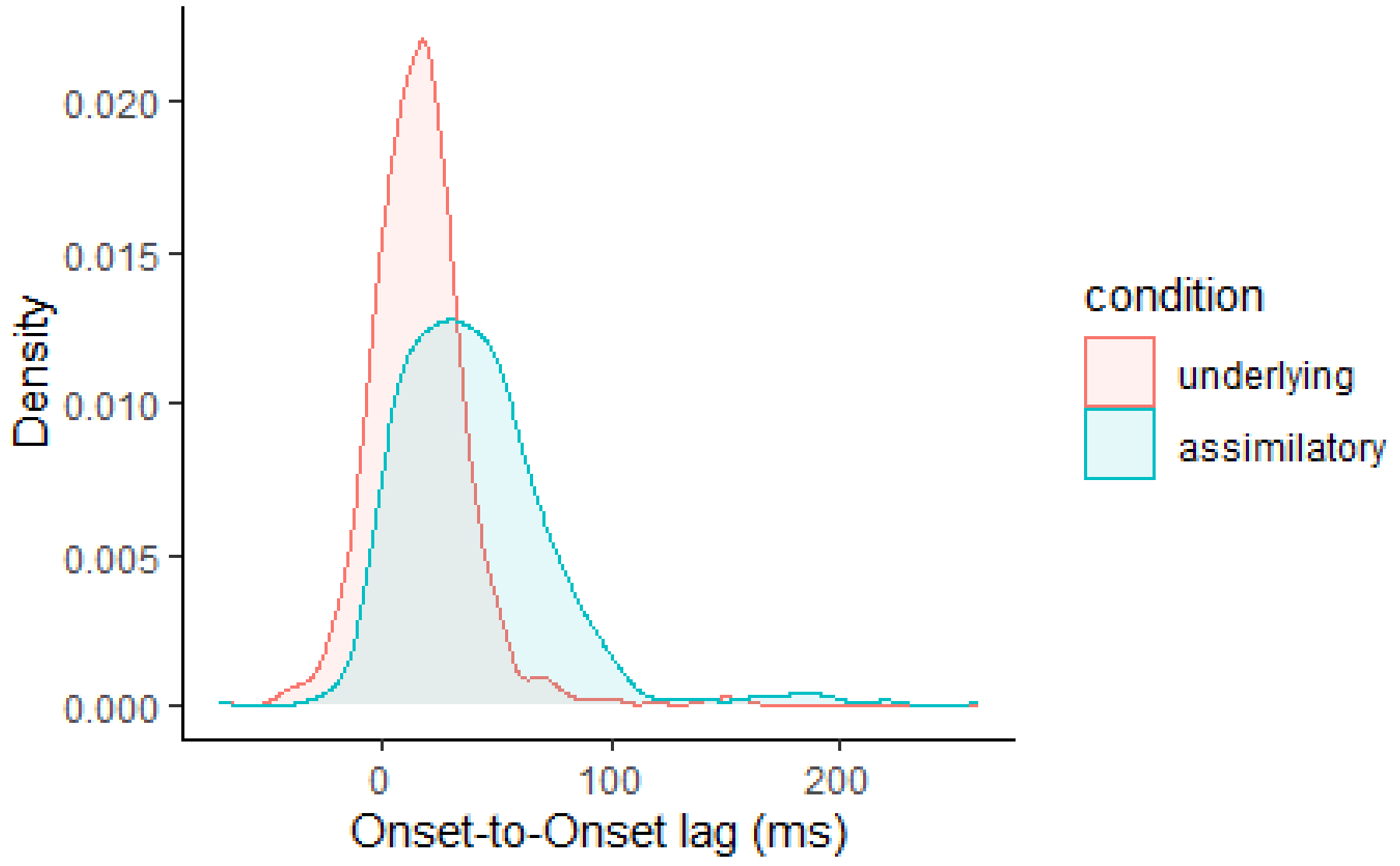

**Figure 8: The distribution of onset-to-onset lag measurements by *Condition*.**

### *4.2. Temporal coordination*

As discussed in Section 2, both the **UNDERLYING** and **ASSIMILATORY** palatalization are expected to show no correlation between *G1 duration* and *onset-to-onset lag*, if both are palatalized. If one (most likely the **ASSIMILATORY** palatalization) turns out to behave like a segment sequence, then the *onset-to-onset lag* will increase with *G1 duration*, leading to a positive correlation between them, as has been observed for English stop-glide sequences (Figure 1).

Figure 9 plots the relation between *G1 duration* (x-axis) and *onset-to-onset lag* (y-axis) across *Condition* for each speaker. To illustrate the trend in the data, a least squares linear regression line is fit to each panel. The $R^2$ value of each regression line from Figure 9 is summarized in Table 2.

The regression line is nearly flat — precisely the pattern predicted for complex segments (Shaw et al., 2021). Notably, this pattern was observed for both **UNDERLYING** and **ASSIMILATORY** palatalization, indicating that plain consonants preceding glides (**ASSIMILATORY** palatalization) are also palatalized. This suggests that the contrast between palatalized and plain consonants is neutralized in this context.

To assess the statistical_significance of the trends in Figure 9, we fit a series of linear mixed-effects models to the data (for additional detail, see Section 3.4). As shown in Table 3, the addition of *Condition* as a fixed factor improves the baseline model, which contains only random effects of subject and item ($\chi^2 = 23.17$, $p < 0.001$). This suggests that the onset-to-onset lag significantly differs by *Condition*, as can be observed in Figure 8 and Figure 9. Crucially, however, the addition of *$G_1$ duration* does not improve the model ($\chi^2 = 2.02$, $p > 0.1$), and neither does the addition of the interaction term ($\chi^2 = 0.86$, $p > 0.1$). The lack of improvement indicates that there is no correlation between *$G_1$ duration* and onset-to-onset lag, which is predicted by complex segment timing.

Table 4 summarizes the best fitting model (onset-to-onset ~ *Condition* + (1 + *$G_1$ duration* | subject) + (1 + *$G_1$ duration* | item)). The main effect of *Condition* is significant ($t = 8.217$, $p < 0.001$). Specifically, **ASSIMILATORY** palatalization is estimated to be 25 ms longer than **UNDERLYING** palatalization in onset-to-onset lag. This pattern is highlighted in Figure 10 which provides a box plot of onset-to-onset lag across *Condition*.

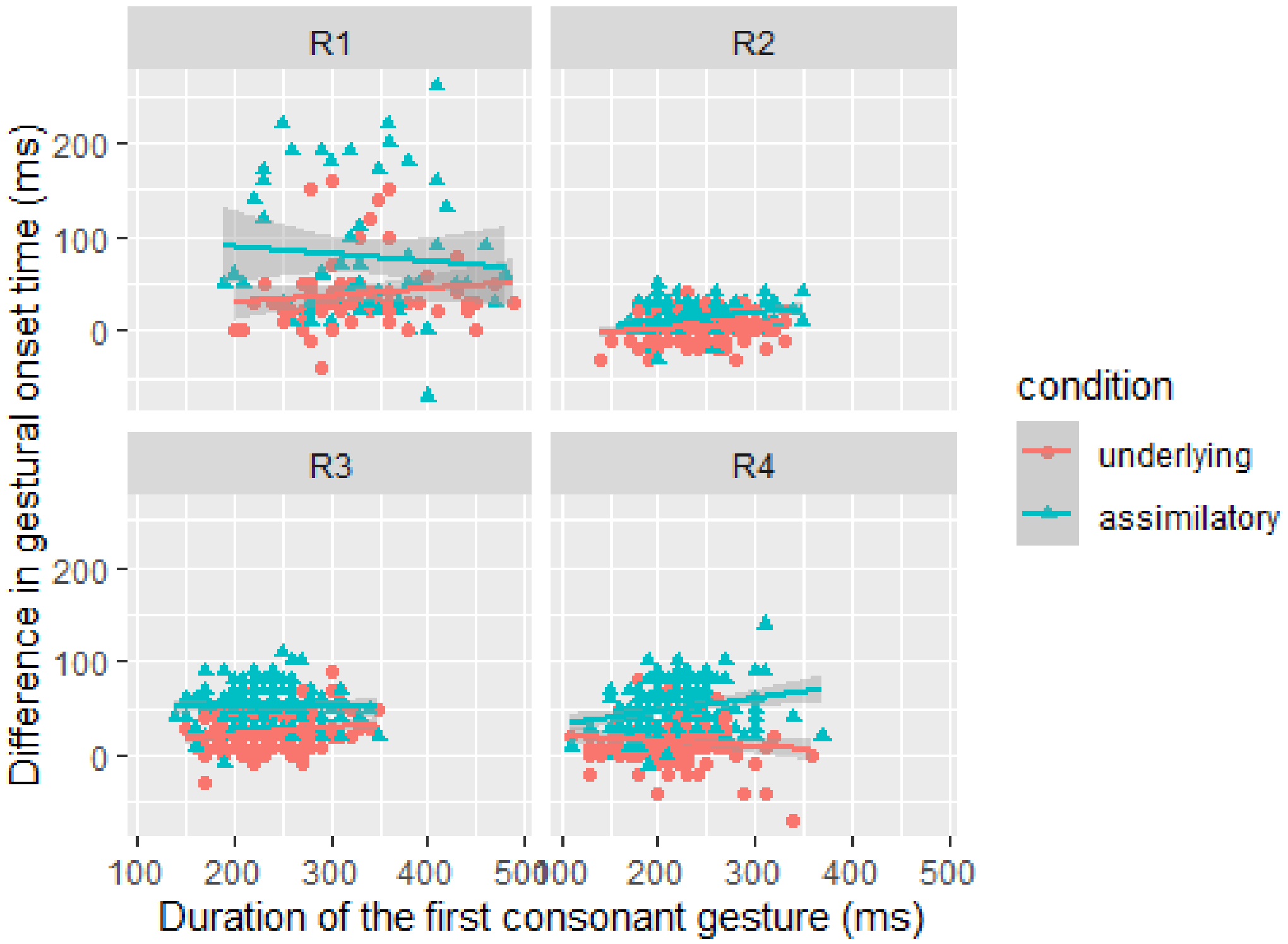

**Figure 9: A scatter plot of the effect of *G1 duration* (x-axis) on *onset-to-onset lag* (y-axis) across *Condition* for each speaker**

**Table 2: Summary of $R^2$ value of each regression line from Figure 9**

| *Condition* | *R1* | *R2* | *R3* | *R4* |
|---|---|---|---|---|
| UNDERLYING | 0.018 | 0.026 | 0.044 | 0.013 |
| ASSIMILATORY | 0.0061 | 0.061 | 0.00001 | 0.052 |

**Table 3: Temporal coordination – Nested model comparison**

| **LME Model comparison (*onset-to-onset*~)** | **Df** | **AIC** | **logLik** | $\chi^2$ | **Pr(>$\chi^2$)** |
|---|---|---|---|---|---|
| 1 + (1 + G1 duration\|subject) + (1+ G1 duration\|item) | 8 | 10511 | -5247.3 | NA | NA |
| 1 + Condition + (1 + G1 duration\|subject) + (1 + G1 duration\|item) | 9 | 10490 | -5235.7 | 23.17 | < 0.001 |
| 1 + Condition + $G_1$ duration + (1 + G1 duration\|subject) + (1 + G1 duration\|item) | 10 | 10490 | -5234.7 | 2.02 | > 0.1 |
| 1 + Condition * $G_1$ duration + (1 + G1 duration\|subject) + (1 + G1 duration\|item) | 10 | 10491 | -5234.3 | 0.86 | > 0.1 |

**Table 4: Temporal coordination – Summary of fixed factors in the best-fitting model (reference level for *Condition*= Underlying)**

| | **Estimate** | **Std.Error** | **df** | **t value** | **Pr(>\|t\|)** |
|---|---|---|---|---|---|
| **(Intercept)** | 32.928 | 8.963 | 8.101 | 3.674 | < 0.01 |
| **Condition** | 25.471 | 3.1 | 8.83 | 8.217 | < 0.001 |

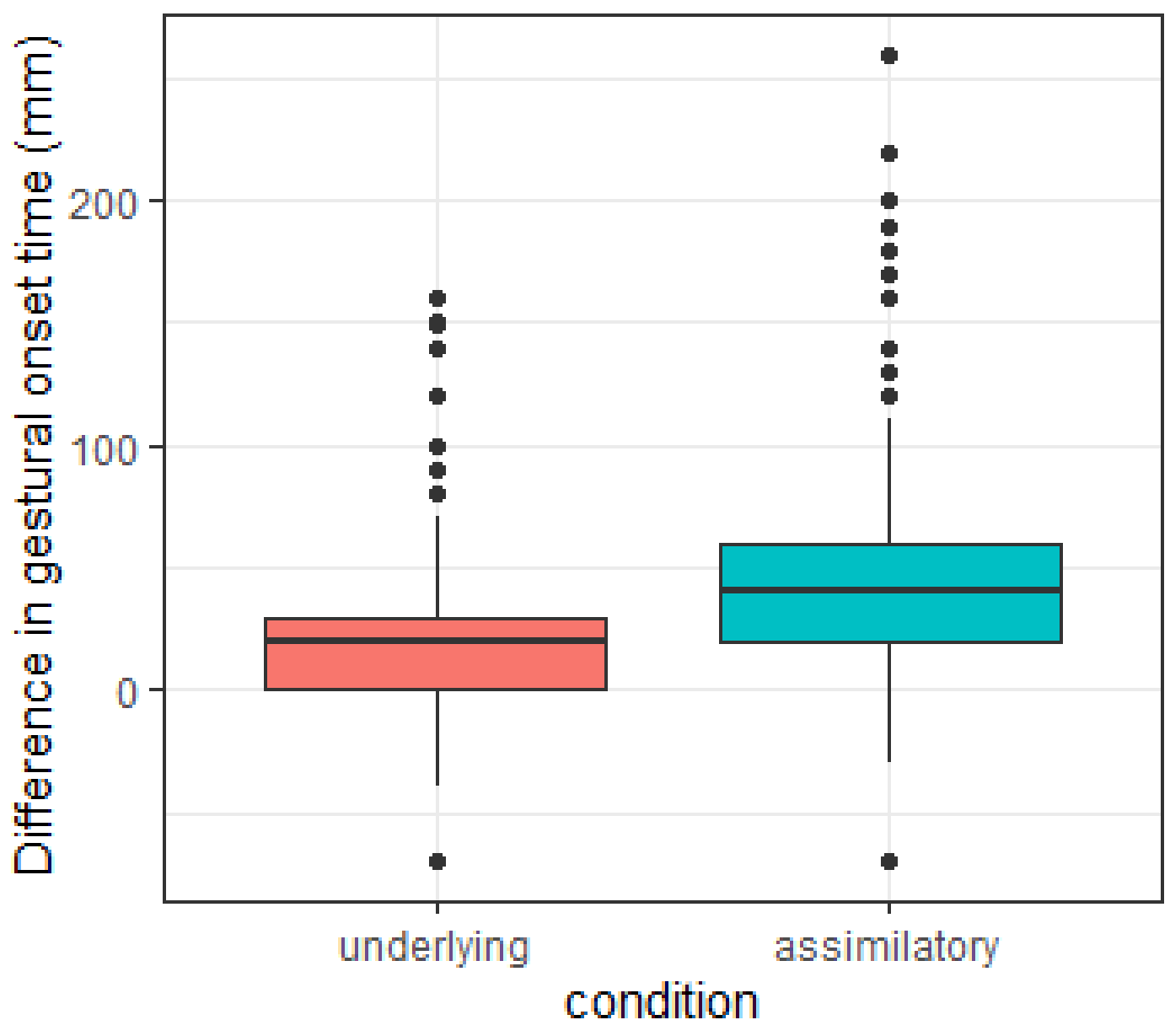

**Figure 10: A boxplot of onset-to-onset lag across *Condition***

In summary, the statistical models confirm the trend observable in Figure 9. There is no clear correlation between G1 duration and onset-to-onset lag, as predicted by the complex segment hypothesis. This contrasts with past work showing that segment sequences show a strong correlation between these measures (see Figure 1). Crucially, the null effect of *G1 duration* indicates that both **UNDERLYING** and **ASSIMILATORY** palatalization have the temporal coordination of complex segments.

### *4.3. Articulatory evidence of incomplete neutralization*

Figure 11 shows the normalized longitudinal position (front-back) of the TB sensors at the gestural onset across conditions. Positive and negative values on the y-axes illustrate the frontness and backness of the tongue body, respectively. The spatial position of the TB is more retracted for the **ASSIMILATORY** palatalization than for the **UNDERLYING** palatalization at the onset of the palatal gesture. As shown in Figure 12, this pattern holds across speakers.

To assess the statistical significance of the trends in Figure 11 and Figure 12, we fit a series of linear mixed-effects models to the data (for additional detail, see Section 3.4). As shown in Table 5, the addition of *Condition* improves the baseline model, which contains only random effects of subject and item ($\chi^2 = 18.846$, $p < 0.001$), indicating that the TB position significantly differs by *Condition*, as observed in Figure 11 and Figure 12. Specifically, the TB is estimated to be 1.5 mm more retracted for the **ASSIMILATORY** palatalization than for the **UNDERLYING** palatalization at the onset of the palatal gesture, as shown in Table 6. This difference is consistent with the presence of a secondary tongue dorsum retraction gesture for plain stops. Some residue of velarization for plain stops persists in the **ASSIMILATORY** condition, in line with the previous observations of an active tongue dorsum retraction gesture in the “plain” stops series.

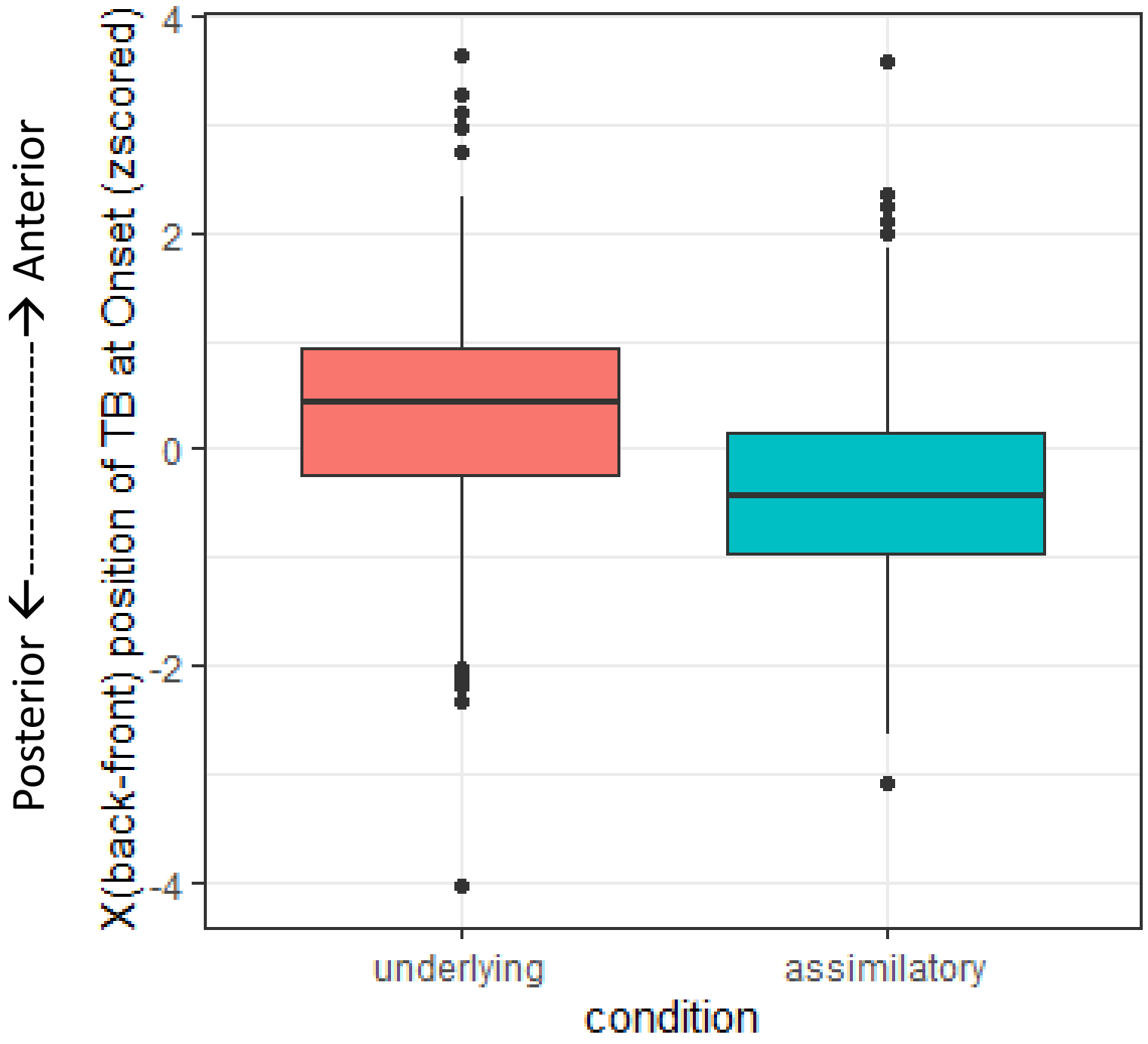

**Figure 11: A boxplot of TB position (z-scored) at palatal gesture onset**

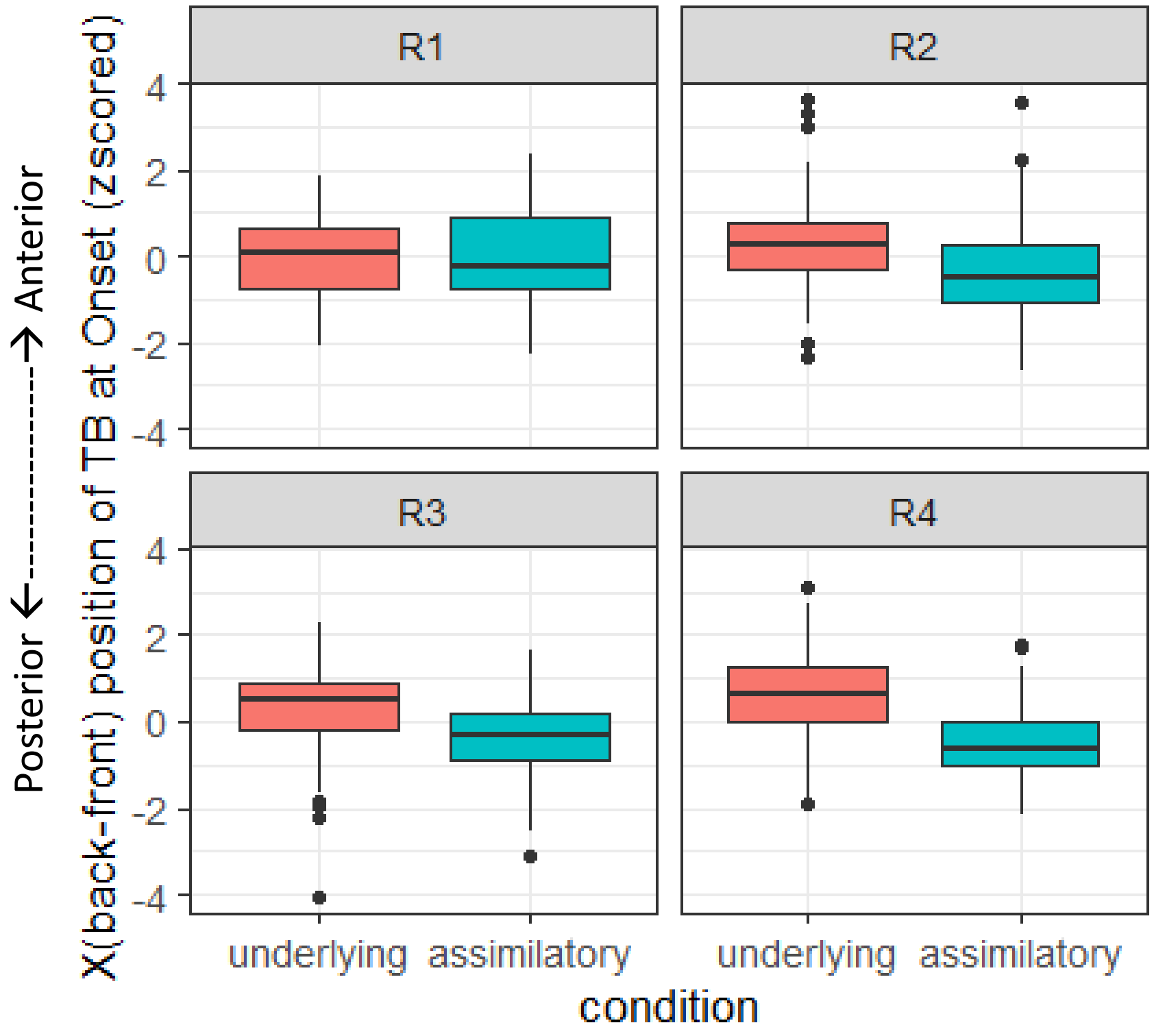

**Figure 12: A boxplot of TB position (z-scored) at palatal gesture onset for each speaker**

**Table 5: TB position – Nested model comparison**

| TB | Df | AIC | logLik | $\chi^2$ | Pr(>$\chi^2$) |
|---|---|---|---|---|---|
| 1 +(1\|speaker) + (1\| sequence) | 4 | 4626.5 | -2309.3 | NA | NA |
| 1+condition +(1\|speaker) + (1\| sequence) | 5 | 4609.7 | -2299.8 | 18.846 | < 0.001 |

**Table 6: TB position – Summary of fixed factors in the best-fitting model (reference level for *Condition*= Underlying)**

| | Estimate | Std.Error | df | t value | Pr(>\|t\|) |
|---|---|---|---|---|---|
| **(Intercept)** | -13.379 | 5.223 | 3.005 | -2.562 | < 0.1 |
| **Condition_Assimilatory** | -1.515 | 0.225 | 10.034 | -6.732 | < 0.001 |

## 5. Discussion

### *5.1. Overview*

Incomplete neutralization has been the focus of much work in laboratory phonology and phonetics. Phenomena which have been described as neutralization have sometimes turned out to show phonetic traces of underlying contrasts. Final devoicing is perhaps the most well-studied case, but there are numerous others, as well as some cases of complete neutralization (see references in the introduction). Determining neutralization status requires careful examination of the phonetic record.

In the present study, we explored a case of putative phonological neutralization, that of palatalized consonants (underlying palatalization; e.g., /b$^j$/) and plain consonants preceding a palatal glide (assimilatory palatalization; e.g., /bj/) in Russian. The purpose of this study is to explore how complete the neutralization is between underlying palatalization and assimilatory palatalization. To do so, we conducted an Electromagnetic Articulography (EMA) experiment and examined the temporal coordination and the spatial positions of articulators involving underlying and assimilatory palatalization in Russian.

We asked two research questions as follows: (1) Do underlying palatalization (e.g., /b$^j$/) and assimilatory palatalization (e.g., /bj/) exhibit temporal coordination characteristic of complex segments? (2) Do underlying palatalization (e.g., /b$^j$/) and assimilatory palatalization (e.g., /bj/) exhibit spatial and/or temporal differences? The first research question was regarding whether two cases of Russian palatalization show neutralization. The second research question was regarding whether the neutralization is complete. Regarding the first question, for there to be evidence of neutralization, both the underlying and assimilatory palatalization should exhibit no correlation between consonant duration and onset-to-onset lag. With regard to the second question, for there to be evidence of incomplete phonetic neutralization, there should be significant spatial and/or temporal differences. Given that plain consonants have secondary velarization (Litvin, 2014; Roon & Whalen, 2019; Skalozub, 1963), we predicted that the gestural blending of two secondary articulation gestures (palatalization and velarization) in assimilatory palatalization would lead to incomplete neutralization of underlying and assimilatory palatalization in Russian. A key finding from the EMA study is that both underlying and assimilatory palatalization types were coordinated as complex segments, according to the hypothesized temporal basis of complex segments (Shaw et al., 2021). Evidence for the coordination pattern came from the relation between G1 duration and onset-to-onset lag. Specifically, these intervals were statistically independent for both underlying and assimilatory palatalization types. The timing of the palatal gesture, G2, was unaffected by the duration of G1, indicating that G2 movement is coordinated with the beginning of G1. The gestures for both underlying and assimilatory palatalization types are coordinated as complex onsets. This suggests that the contrast between a palatalized consonant and a plain consonant is neutralized to the palatal counterpart when a plain consonant is followed by a glide.

However, underlying and assimilatory palatalization do show significant phonetic differences in other dimensions. The onset lag is longer for the assimilatory palatalization than for the underlying palatalization. Furthermore, the spatial position of the articulators also provided residual evidence of an underlying tongue dorsum retraction for assimilatory palatalization. In particular, the spatial position of the TB at the onset of the palatal gesture was more retracted for the assimilatory palatalization condition than for the underlying palatalization condition. This is in line with previous findings of Russian plain consonants having secondary velarization, and may in fact be underlyingly velarized, /Cˠ/ (Padgett, 2001; Rubach, 2000). As predicted in Section 2, the gestural overlap on the same tract variable (i.e., palatalization vs. velarization) would lead to gestural blending between these two gestures. Accordingly, this results in a slightly more retracted tongue position for the assimilatory palatalization compared to underlying palatalization, which only has the palatal gesture on the TB tract. Consequently, this difference leads to incomplete neutralization between the underlying and assimilatory palatalization in Russian.

Although we focused on the C+j sequence to examine assimilatory palatalization, this case is part of a larger process of C + palatal(ized) segments in Russian. That is, “plain” consonants tend to undergo palatalization when followed by any palatal or palatalized consonant (e.g., Timberlake, 2004). For example, /s/ in a word ‘campfire’ /kˠoˈsˠtʲorˠ/ is also realized as [sʲ], resulting in [kˠʌˈsʲtʲorˠ] (cf. [kˠʌˈsˠtˠrˠɨ] ‘campfire (pl)’). Furthermore, though assimilatory palatalization in C+Cʲ sequences is beyond the empirical scope of this study, we predict that the target consonant in such cases would phonetically behave similarly to the target consonant in C+j sequences. For example, if we examine /s/ preceding /tʲ/ (C+Cʲ sequence of assimilatory palatalization) and compare it with underlying palatalization e.g., /sʲ/, we also expect incomplete neutralization between them. That is, /sˠ̠/ preceding /tʲ/ will also exhibit the temporal coordination of complex segments, while showing residual tongue dorsal retraction.

### *5.2. Incomplete neutralization as gestural overlap*

From the EMA data, we found residual evidence of an underlying tongue dorsum retraction in the assimilatory palatalization condition. This finding is in line with the prediction made in Section 2. That is, the gestural overlap resulting in gestural blending of two secondary articulation gestures (palatalization and velarization) leads to incomplete neutralization of underlying and assimilatory palatalization in Russian. This process also fits in with other empirical cases, such as vowel assimilation in Igbo (Zsiga, 1997) and blended vowels in Romanian (Marin, 2012).

In Igbo, when a vowel is followed by another vowel, the preceding vowel is assimilated to the following vowel (i.e., complete assimilation: V1V2 --> V2V2). For example, /e/ in /nwoke a/ ‘this man’ is assimilated to the following vowel /a/ (i.e., /nwoke#a/ --> [nwoka#a]). However, previous studies have reported such vowel assimilation exhibits gradient behavior (Clark, 1990; Emenanjo, 1978; Zsiga, 1997). For example, for /nwoke#a/, the realization of /e/ is different from a typical vowel realization, as the realization of /e/ varies from more [e]-like realizations to more [a]-like realizations. Zsiga (1997) argued that this is due to the shortening of V1 in the word final position and compensational lengthening of V2. Zsiga rejected the gestural blending account "because it fails to link assimilation to final reduction." However, it may be possible to link assimilation to final reduction in the AP framework, if we assume that weaker blending strength contributes to reduction. When V1 occurs at the word-final position, where vowels are reduced, the blending strength of V1 is also reduced. This proposal involves a dependency between blending strength and word position. Introducing this type of dependency makes it possible to describe Igbo vowel assimilation in terms of blending between two gestures V1 and V2, with varying different degrees of overlap between the two gestures. For example, for /nwoke#a/, the blending strength of /e/ is weaker than that of /a/, as /e/ occurs at the word-final position. When there is less or no overlap between two vowels, /e/ is realized more like [e]. On the other hand, when the two vowels significantly overlap, the realization is more like [a], as the blending strength for /a/ is stronger. However, the [a]-like realization is expected to be different from a typical /a/ vowel, as it comes from the blending of two vowels.

Marin (2012) also proposed a production model to examine the incomplete neutralization of Romanian vowels. In Romanian, the vowel [e] alternates with diphthong [ea] (derived [e]); acoustic analysis revealed that for Romanian vowels, the derived [e] is significantly more central than the vowel [e] that is underlyingly /e/ (underived [e]). She hypothesized that it might be attributable to different production mechanisms between derived and underived [e], and tested her hypothesis by comparing acoustic data to modeled stimuli. Using an articulatory based synthesizer, *TADA* (*Task Dynamic Application*; see e.g., Nam et al., 2006), the underived /e/ was modeled with the gestural specifications of a single gesture [e], while the derived /e/ was modeled as the blending of two gestures [e] and [a], reflecting its underlying status as diphthong, /ea/. The results revealed that the blending of two gestures /e/ and /a/ showed similar acoustic properties to naturally produced derived [e], and modeled stimuli for underived [e] were also similar to naturally produced underived [e]. Along with our findings, Marin's results also support that at least some cases of incomplete neutralization can be modeled as gestural overlap.

### *5.3. Alternative account: the effect of back vowels on tongue body position*

The present study has demonstrated that incomplete neutralization might be attributable to two secondary articulation gestures (palatalization and velarization) in assimilatory palatalization. However, there is an alternative explanation for the retracted tongue position observed as a result of this process. Namely, it is possible that the retracted tongue position is attributable to the blending of the palatal gesture and the following vowel gesture.

Consonant-to-vowel coarticulation is commonly found crosslinguistically, such as in English (e.g., Keating, 1993), Russian (e.g., Iskarous & Kavitskaya, 2010), French (e.g., Guitard-Ivent et al., 2021), Catalan (e.g., Recasens, 1985), and Algerian Arabic (e.g., Bouferroum & Boudraa, 2015). Given that the target vowels are all back vowels (/u/ and /o/) in the experiment, the retracted tongue position of the assimilatory palatalization could be the result of blending the palatal gesture and the back vowel gesture. For this to be a viable alternative, we would have to consider as well how the vowel is coordinated with the other gestures in our target items. For predictions about gestural phasing patterns between consonants and vowels, we turn to the coupled oscillator model of syllable structure.

The coupled oscillator model seeks to explain the observations that syllable structure is associated with a characteristic pattern of temporal coordination (Goldstein et al., 2006; Goldstein et al., 2009; Nam et al., 2009; Saltzman et al., 2008). To explain observed overlap between onset consonants with the following vowel, it is hypothesized that a gesture in a syllable onset is coordinated in-phase with the following vowel. The implication is that the two gestures are triggered at the same time. In contrast, a coda gesture is hypothesized to be coordinated anti-phase with the preceding vowel, showing a sequential timing between the two gestures. Furthermore, the coupled oscillator model hypothesizes that multiple gestures in a syllable onset are coupled anti-phase with each other, along with both being in-phase with the vowel. To satisfy this competitive coupling demand, one gesture (C1) shifts away from the vowel, and the other gesture (C2) shifts toward the vowel. Consequently, for gestures in the syllable onset position, the midpoint of prevocalic consonants exhibits a stable timing with the following vowel, regardless of the number of onset consonants. This pattern, the so-called “c-center effect”, has been observed in several studies (Browman & Goldstein, 1988; 2000; Crouch, Katsika, & Chitoran, 2020; 2022; Goldstein et al., 2007; Goldstein, Chitoran, & Selkirk, 2007; Marin, 2013; Marin & Pouplier, 2010; Nam & Saltzman, 2003; Shaw & Gafos, 2015; Sotiropoulou & Gafos, 2022).[iv]

The differences in TB position observed across palatalization conditions, which we attributed to incomplete neutralization, might instead be attributable to a different vowel coordination pattern between underlying and assimilatory palatalization. For underlying palatalization (e.g., /bʲusʸ̠tʸ̠/ [bʲusʸ̠tʸ̠]), the labial and palatal gestures might be coupled in-phase with each other and also coupled in-phase with the vowel. Due to these coupling relations, the labial and palatal gestures will start at the same time with the following vowel. In contrast, for assimilatory palatalization (e.g., /bʸ̠jutʸ̠/ [bʲjutʸ̠]), the labial and palatal gestures might be competitively coupled with the vowel, i.e., coupled anti-phase with each other and also coupled in-phase with the following vowel. Due to this competitive coupling, the vowel gesture for /u/ would start before the palatal gesture and continue concurrently during the palatal gesture. In both cases, the temporal overlap between the palatal gesture and the following vowel gesture in the same tract variable (TB) will lead to gestural blending between them. Crucially, however, when the spatial position of the TB gestures is compared at the onset of the palatal gesture, there may be differences, owing to the differences in temporal coordination. Assimilatory palatalization, which has the tongue backing for /u/ starting earlier than the palatal gesture, will show a more retracted tongue position than the underlying palatalization condition, since the backing movement for /u/ will have already started before the palatal gesture starts, just in the assimilatory palatalization condition. That is, due to the existence of the vowel gesture preceding the palatal gesture for the assimilatory palatalization, the gestural blending of the tongue backing for /u/ and the fronting for /j/ may result in a more retracted tongue position at the onset of the TB gesture for the assimilatory palatalization as compared to the tongue position for the underlying palatalization. This is a reasonable alternative that is compatible with both the temporal basis of complex segments and the coupled oscillator model of syllable structure.

Other aspects of our data serve to rule out this alternative hypothesis. Crucially, if the labial gesture and the palatal gesture are coordinated anti-phase in the assimilatory palatalization condition, similar to segment sequences in English such as /bjut/ ‘butte’ (Shaw et al., 2019; Shaw et al., 2021), they are expected to result in the temporal coordination of segment sequences, which is not the case in our data (Figure 9). Consequently, the back vowel is less likely to be the source of incomplete neutralization. C-center timing treats the consonantal gestures as separate segments (timed anti-phase) and is therefore incompatible with complex segment coordination. Thus, while the difference in tongue position at the onset of the palatal gesture is consistent with c-center timing, the data on the whole is not.

### *5.4. Does the delayed onset of the palatal movement also follow from blending?*

As discussed above, gestural blending between palatalization and velarization in the assimilatory palatalization condition may lead to a more retracted tongue position than would be expected for underlying palatalization. However, it is not clear what causes the delayed onset of the tongue body movement associated with the palatal gesture in the assimilatory palatalization condition. Here, we consider whether gestural blending could indirectly condition this difference, or, alternatively, whether some other parameter is responsible. In all we consider three possible explanations.

The most parsimonious explanation would be if the gestural blending parameters posited to explain the spatial differences across conditions also account for the temporal difference. In the AP framework, the blending of the dynamical parameters of two gestures is predicted to produce an outcome that falls somewhere in-between the two gestures, depending on the strength of the two gestures in question (e.g., Browman & Goldstein, 1989; 1992). We first consider the possibility that the tongue body gesture starts at the same time for both palatalization conditions, but that blending delays the detection of the movement onset in the assimilatory palatalization condition. We explored this possibility through simulation in TADA; however, we found that varying the gestural blending strength alone was insufficient to derive differences in onset lag. Detection of gestural onsets, using our methods (see Section 3.4) is sensitive to the stiffness of the gestures but not to blending strength. Thus, at least in the version of AP implemented in TADA, blending strength alone cannot account for the difference in onset lag across conditions.

A second possible explanation is to posit that, in the assimilatory palatalization condition, the labial and palatal gestures are coordinated anti-phase, and the labial and velar gestures are coordinated in-phase. In the underlying palatalization condition, by contrast, the labial and palatal gestures are coordinated in-phase. Such a coordination difference will result in the delayed onset of the TB gesture in the assimilatory palatalization condition relative to the underlying palatalization condition. The delay of the TB gesture will generate a delay in movement onset, similar to that observed in our data. However, this leads to the same problem as the competitive coupling account discussed above in Section 5.3. The anti-phase coordination between the gestures for /b/ and /j/ is expected to produce a positive correlation between C1 and onset-to-onset interval, the temporal coordination pattern characteristic of segment sequences. As we noted, this is not the case for assimilatory palatalization in Russian. The increased onset-to-onset lag emerges alongside evidence for complex segment (as opposed to segment sequence) coordination.

Lastly, we consider the possibility that in assimilatory palatalization the velar gesture starts before the palatal gesture and continues concurrently. On this proposal, the labial and palatal gestures are still coordinated in-phase, the coordination relation for complex segments. This gets around the problems of other approaches by successfully deriving the observed relation between C1 duration and onset-to-onset lag (Figure 9). However, in order to derive the differences across conditions, the velar gesture is eccentrically timed to the palatal gesture so as to start slightly earlier in time. Eccentric timing refers to phasing relations that are neither in-phase nor anti-phase and have been argued to be empirically necessary in some languages (e.g., Goldstein, 2011; Geissler et al., 2021). In this case, eccentric timing can derive delayed onset of the TB movement for the assimilatory palatalization while maintaining the temporal coordination of complex segments. Of the three theoretical possibilities, the first two are more parsimonious. However, the only set of gestural parameters that derives the complete range of empirical facts involves both blending of velar and palatal gestures and an eccentric timing relation between them.

## 5.5. *Conclusion*

Russian contrasts palatalized and non-palatalized (plain or velarized) consonants, but this contrast is reported to be neutralized when a plain consonant is followed by a palatal glide or a palatal consonant (e.g., Avanesov, 1972; Timberlake, 2004). In this paper, we used Electromagnetic Articulography (EMA) to explore the neutralization of palatalized consonants (underlying palatalization; e.g., /b$^{j}$/) and velarized consonants preceding a palatal glide (assimilatory palatalization; e.g., /b$^{ɣ}$j/).

A key finding from the EMA experiment is that both underlying and assimilatory palatalization exhibited temporal coordination characteristic of complex segments, showing no correlation between consonant duration and onset-to-onset lag. This suggests that the contrast between a palatalized consonant and a velarized consonant is neutralized to the palatal counterpart when a velarized consonant is followed by a glide. However, the neutralization of the secondary articulation contrast is phonetically incomplete. In particular, we found that the tongue body was significantly more retracted for the assimilatory palatalization than for the underlying palatalization at the onset of the palatal gesture. The difference across conditions was small (1.5 mm) but consistent across speakers. In addition, *Onset-to-onset lag* is significantly longer (25 ms) for assimilatory palatalization than for underlying palatalization. These significant differences suggest that the neutralization of the palatalized-velarized contrast is phonetically incomplete. Furthermore, the residual evidence of an underlying tongue dorsum retraction for the assimilatory palatalization is in line with previous findings of Russian non-palatalized consonants having secondary velarization.

Incomplete neutralization has been argued to be a serious challenge to models of the phonology-phonetics interface, particularly those that deal with the neutralization component in terms of symbol substitution. The current study offers an explanation for incomplete neutralization patterns by showing that at least some cases of incomplete neutralization can be modeled as gestural overlap and blending in the Articulatory Phonology framework. There is substantial potential for the gestural overlap account to generalize across a wide range of incomplete neutralization cases.

---

[i] Palatalization in C + /j/ sequences is part of a larger assimilatory process of C + $C^j$ sequences, the application of which depends on a range of factors, including the place and manner of the consonants, presence of morpheme boundaries, stress, etc. (Avanesov, 1972; Timberlake, 2004).

[ii] The Korean results are not discussed as they are not relevant to the current paper.

[iii] In the verbs with the /Cj-/ sequence in Table 1, the glide is derived from the vowel /i/ (e.g., /pj-o-t/ 'drink (3ps pres)' from /pi-$t^j$/ 'drink (inf)').

[iv] Not all languages with complex syllable onsets show the c-center effect. Brunner et al. (2014) show that some German clusters do not show the c-center effect, although they show other indices of consistent with global timing (Sotiropoulou & Gafos 2022). Georgian consonant clusters also do not show the c-center effect, possibly due to reasons related to the morphology of the language (Crouch et al., 2020, 2022, c.f., Goldstein et al. 2007).